\documentstyle [floats,prl,aps]{revtex}
\input epsf
\begin{document}

\draft
\title{Possible low-background applications of MICROMEGAS detector technology}

\author{
J.I. Collar$^{a,b}$ and Y. Giomataris $^{c}$
}
\address{ 
$^{a}$Groupe de Physique des Solides (UMR CNRS 75-88), Universit\'es Paris 
7 \& 6, 
75251 Paris Cedex 05, France\\
$^{b}$Centro de F\'\i sica Nuclear, Universidade de Lisboa, 1649-003 Lisbon, 
Portugal\\
$^{c}$CEA/Saclay, DAPNIA, 91191 Gif-sur-Yvette Cedex, France\\
}
\wideabs{
\maketitle
\begin{abstract}
\widetext
The MICROMEGAS detector concept, generally optimized for use in 
accelerator experiments, displays a peculiar combination of features that can 
be advantageous in several astroparticle and neutrino physics applications. 
Their sub-keV ionization energy threshold, excellent energy and space resolution, 
and a simplicity of design that 
allows the use of radioclean materials in their construction are some of these 
characteristics. We envision tackling experimental challenges such as the 
measurement of neutral-current neutrino-nucleus coherent scattering or Weakly 
Interacting Massive Particle (WIMP) detectors with directional sensitivity. 
The large physics potential of a compact (total volume 
O(1)m$^{3}$), multi-purpose array of 
low-background MICROMEGAS is made evident.
\end{abstract}

\pacs{ {\tt  Presented at IMAGING 2000, Stockholm, June 2000}\\
}}
\narrowtext
{\bf 1. MICROMEGAS operating principle.}

\vspace{1mm}
MICROMEGAS (MICROMEsh GAseous Structure, $\mu$MS here) is a 
two-stage parallel-plate avalanche chamber design consisting of a 
100 $\mu$m narrow amplification gap and a thick (up to several cm) conversion 
region, separated by a gauze-light electroformed conducting micromesh. Electrons released 
in the gas-filled conversion gap by an ionizing particle are drifted to the 
amplification gap where they are multiplied in an avalanche process. 
Detectable signals are then induced on anode elements (strips or pads). These 
provide an accurate monodimensional or x-y spatial projection of the energy 
deposition. The distance between the micromesh (cathode) grid and the anode 
plane is maintained by spacers 150 $\mu$m in diameter, placed every 
$\sim$1 mm.  
These are printed on a thin substrate by conventional lithography of a 
photoresistive polyamide film. The anode plane is similarly printed on a thin 
epoxy or Kapton foil. Further details of the $\mu$MS design can be found in 
\cite{mms1,mms2,mms3,mms4}. The extreme rate capability of these 
devices 
($\sim$10$^{6}$ particles mm$^{-2}$ s$^{-1}$) 
allows for their use in medical X-ray imaging.
Their present maximum space resolution 
of O(10) $\mu$m \cite{mms4} gives them 
an obvious utility in high energy physics as tracking chambers. 
Here we concentrate on the alternative 
possibility of exploiting some of the features specific to the $\mu$MS in a different 
realm, that of searches for rare-events in neutrino and astroparticle physics. 

\vspace{1mm}
{\bf 2. Detector features of interest.}

\vspace{1mm}
A common denominator to the applications proposed here is the need to 
extrude a sporadic signal from a mound of competing backgrounds. Several 
properties of $\mu$MS  work together toward this goal:

\vspace{1mm}
$\bullet$  Their excellent spatial resolution
allows for precise particle 
identification based on range {\frenchspacing vs.} energy considerations. The 
intrinsic time resolution of the detector is below 1 ns and therefore an 
accurate third dimension can be extracted from time measurements, allowing 
for complete track reconstruction when necessary.
\begin{figure}[tbp]
\epsfxsize = \hsize \epsfbox{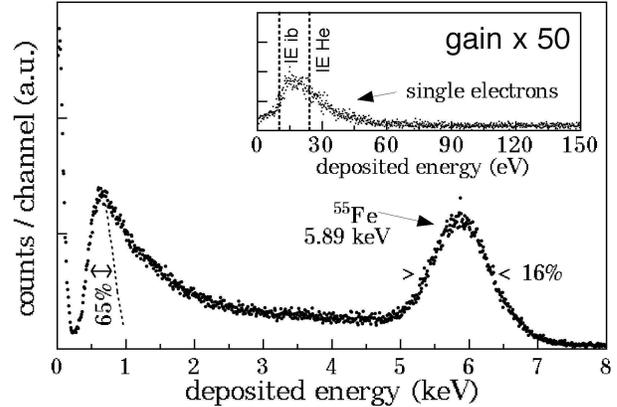}
\caption{Response 
of an unshielded  He + 10$\%$  iC$_{4}$H$_{10}$ $\mu$MS 
chamber ($10\!\times\! 10\!\times\! 1.5$ cm$^{3}$ conversion volume) to a weak
$^{55}$Fe source at 1 bar. 
The assymetric signal at low energy (present only at non-zero drift 
field) is presumably due to cosmic muons of $dE/dx\sim 0.4$ keV/cm. 
{\it Inset:} Blow-up of the threshold 
region: the peak at $\sim$20 eV is probably due to single-electron
field emission from the nickel micromesh.  The frequency of this process
(here 10$^{-2}$ Hz)  
strongly depends on operating conditions and was not optimized during this run. 
Ionization energies for isobutane and He 
are indicated, evidencing the good linearity in energy of the device.} 
\end{figure}

\vspace{1mm}
$\bullet$ Very often the signal sought falls in the keV or sub-keV energy region, as is 
the case in solar axion or WIMP searches. The high gains (up to 10$^{5}$-10$^{6}$) obtained 
with these chambers permit to detect single electrons with $\sim\!100\%$ efficiency 
\cite{mms5,philippe}. This translates into 
an effective energy threshold close to the ionization energy of the gas, i.e., few 
tens of electronvolts (\frenchspacing{Fig. 1}). 
Having been conceived as a TPC detector, $\mu$MS is 
compatible with large drift volumes and operation at high pressure, an 
example of which are the HELLAZ prototypes \cite{philippe}. Large $\mu$MS 
detectors (40 cm $\!\times\!$ 40 cm $\!\times\!$  3 mm) 
are presently in use \cite{thers}. A 
modular design providing the required performance specifications for the 
searches discussed here and a total target mass of several kg within a 
compact O(1)m$^{3}$ volume is feasible using standard $\mu$MS technology. Until 
now, the combination of substantial target mass (an obvious must in rare-event
searches) and sub-keV energy threshold constituted a void in detector 
technology.

\vspace{1mm}
$\bullet$ A typical $\mu$MS energy resolution already approaches that of conventional 
semiconductor detectors in this low energy region, a remarkable feat for gas 
devices. A 5.4$\%$ keV FWHM at 22 keV has been recently obtained 
\cite{delbart} (\frenchspacing{Fig. 2}). This is required for positive identification of signals 
that extend over just a few keV.
\begin{figure}[tbp]
\epsfxsize = \hsize \epsfbox{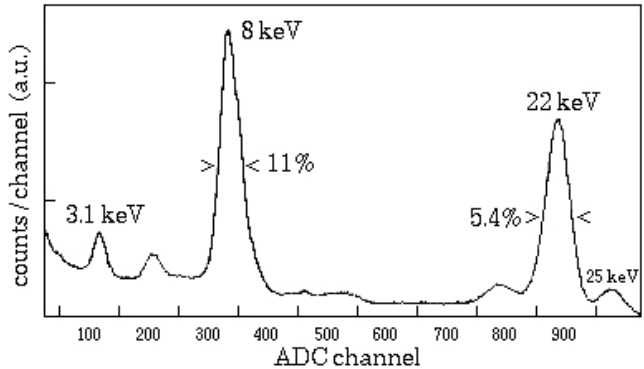}
\caption{Response 
of an Ar + 10$\%$  iC$_{4}$H$_{10}$ $\mu$MS chamber to 
collimated $^{109}$Cd X-rays at 1 bar. The energy resolution is already comparable
to that of large Germanium detectors [39].} 
\end{figure}

\vspace{1mm}
$\bullet$ A large variety of target gas mixtures is available (He, Ne, Ar, 
Xe, CF$_{4}$, 
\frenchspacing{etc.}) and compatible 
with high gain operation. This versatility in atomic mass and gas 
density can be used for 
signal and/or background characterization. For instance, the ultimate 
limitation to WIMP searches is the background due to nuclear recoils from 
neutron scattering \cite{edelweiss}, identical to those expected from WIMP interactions. 
However, the dependence on target nucleus differs from one to the 
other.

\vspace{1mm}
$\bullet$ The structural simplicity and ease of assembly 
of these chambers allows to manufacture them 
out of radioclean materials without much ado. The only components essential 
to the $\mu$MS design are all in the form of very thin foils, with a large freedom 
in the choice of building materials. Besides, intrinsic backgrounds such as 
spurious pulses, common in other high-gain gas detectors such as the 
Multi-Wire Proportional Chamber (MWPC), are comparatively less 
important in $\mu$MS.

\vspace{1mm}
$\bullet$ Last but not least, their fabrication cost is low enough ($\sim$1 
USD/cm$^{2}$) to 
envision moderate budgets for the projects discussed next.

\vspace{1mm}
{\bf 3. Applications.}

\vspace{1mm}
{\bf 3.1 Solar axion searches.}

\vspace{1mm}
The axion, an as-of-yet hypothetical particle 
arising from the Peccei-Quinn broken symmetry \cite{peccei},
has been the object of an extensive 
theoretical and experimental activity \cite{axions}. If axions exist, they can 
play an important role in stellar dynamics. This has prompted several 
searches for axion emission from our Sun \cite{axions}. 
Their energy spectrum, a reflection of the inner solar temperature, 
is expected to peak in the few keV region, dying off at $\sim$10 keV. A recently approved 
experiment \cite{satan} at CERN aims to revisit the ``axion helioscope''
approach, where an attempt is made to 
convert solar axions to detectable 
low-energy photons by use of a strong magnetic field, provided there with 
unprecedented intensity by a decommissioned LHC test magnet. This 
experiment will be performed above ground due to the large size of the 
cryogenic magnet. As a result, background rejection will be essential in 
achieving the intended sensitivity. The use of $\mu$MS chambers can largely reduce 
the expected background \cite{satbac} which in this case will mostly
consist of minimum ionizing particles leaving partial energy depositions in 
the spectral region of interest but having ranges that are too long to originate 
from an axion-induced photoelectron, or neutron-induced recoils of similar 
energies having ranges that are too short. 
The low energy threshold will 
allow to resolve the full axion spectrum, which in some models can peak at 
energies as low as 0.8 keV. 
Large $\mu$MS of a dimension adequate for this search ($\sim10$ cm 
drift length) have been tested 
\cite{sattes} and prototypes built with radioclean materials are under 
construction. Other projects discussed in this section should profit from this 
R$\&$D.

\vspace{1mm}
{\bf 3.2 Coherent neutral-current neutrino-nucleus scattering.}

\vspace{1mm}
An uncontroversial process in the Standard Model, the scattering off 
nuclei of low-energy neutrinos ($< $ few tens of MeV) via the neutral current 
\cite{freedman} remains undetected. The long neutrino wavelength probes 
the entire nucleus, giving rise to a large coherent enhancement in the cross 
section, proportional to neutron number squared \cite{drukier}. 
A quantum-mechanical condition for the appearance of coherent effects is the 
indistinguishability of initial and final states and hence the absence of a 
charged-current equivalent. In principle, it would be possible to speak of 
{\it portable} neutrino detectors since the expected rates can be as high as several 
hundred recoils/kg/day (table 1), by no means a ``rare-event'' situation. 
However, the recoil energy transferred to the target is of a few keV at most for 
the lightest nuclei (\frenchspacing{Fig. 3}). Of this, only 
a few percent goes into ionization, 
most of it being lost as phonons. This ionization energy is generally referred to as 
``electron equivalent energy''  (e.e.e.) 
and its fraction of the total as ``quenching 
factor''.  

The interest in observing this process is not merely academic: a 
neutral-current detector responds the same way to all know neutrino types.
Therefore, the observation of neutrino oscillations in such a device would 
be direct evidence for a fourth sterile neutrino. These must be invoked if all 
recently observed neutrino anomalies are accepted at face value 
\cite{john} and may play an important role as dark matter \cite{dolgov}. 
Separately, the cross section for this process is critically dependent on 
neutrino magnetic moment: concordance with the Standard Model prediction 
would {\it per se} largely improve the present experimental sensitivity 
to $\mu_{\nu}$\cite{dodd}. 
Finally, this coherent mechanism plays a most important role in neutrino 
dynamics in supernovae and neutron stars \cite{freedman}, adding to the 
attraction of a laboratory measurement.
\begin{table}
\caption{Expected number of neutral-current nuclear recoils in several 
MICROMEGAS gases from a typical reactor neutrino flux and spectrum 
(10$^{13} \nu$ cm$^{-2}$ s$^{-1}$).}
\begin{tabular}{c c c c}
\bf{Gas} & \bf{ Recoils / kg / day} & \bf{of which} & 
\bf{of which}\\
 &  & \bf{E$_{rec}<$1 keV} & 
\bf{E$_{rec}<$100 eV}\\
He & 8.1 & 3.5  & 0.6\\
Ne & 43.9 & 37.0 & 10.7\\
Ar & 103.2 & 98.1 & 39.8\\
Xe & 389.7 & 389.1 & 274.1\\
\end{tabular}
\end{table}

Several proposals to use phonon-sensitive new-generation cryogenic 
detectors \cite{blas,starostin} have been put forward, but no existing device 
meets the mass and energy threshold requirements involved in this 
measurement. The advent of the $\mu$MS concept may have broken this impasse: 
\frenchspacing{Fig. 3} shows the signal expected in a reactor 
experiment for several gases 
used in $\mu$MS. A considerable fraction of the signal is found in all cases above the 
achieved $\mu$MS thresholds. Experimental quenching factors are nevertheless 
{\it terra incognita} at these low energies (here they have been extrapolated below 
1 keV recoil energy from SRIM stopping power tables \cite{srim}). It is 
therefore imperative to perform preliminary measurements with a well-characterized 
radiation source able to produce {\it exclusively} these low-energy 
recoils. Adequate monochromatic (filtered) neutron beams 
\cite{nbeams} with negligible photon contamination 
are available from a handful of research reactors.  
The expected recoil 
signal falls in the same energy region as for reactor neutrinos 
(\frenchspacing{Fig. 3}, inset). 
These preliminary proof-of-principle measurements could be performed 
immediately with the HELLAZ prototype chamber \cite{philippe2}. Indeed, 
this large-volume (25 $ l$), high-pressure (20 bar) $\mu$MS prototype seems ready 
for a first-ever measurement of this interesting mode of neutrino interaction.
\begin{figure}[tbp]
\epsfxsize = \hsize \epsfbox{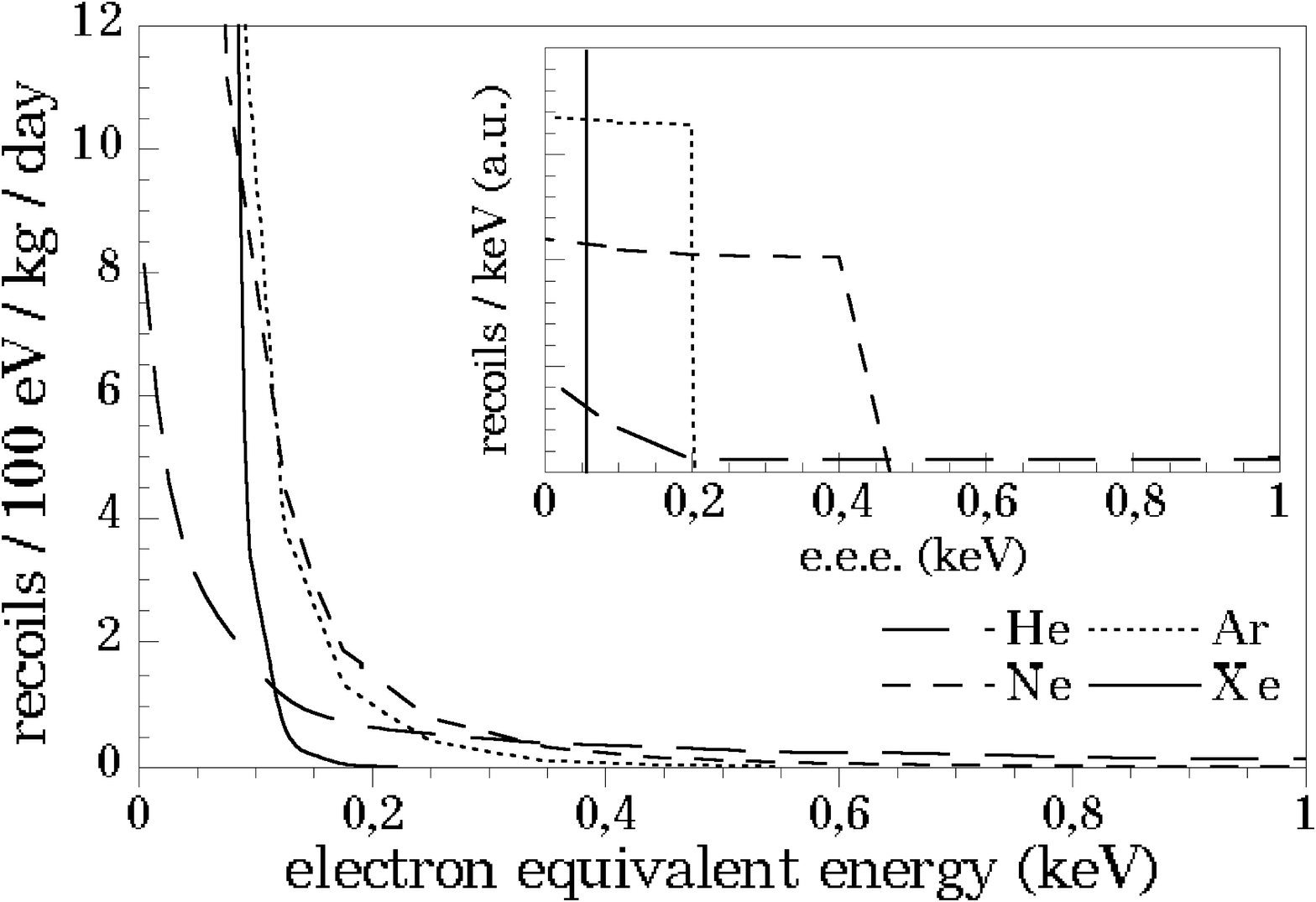}
\caption{Detectable signal in different MICROMEGAS 
gases from neutral-current nuclear scattering of reactor 
neutrinos (10$^{13} \nu$ cm$^{-2}$ s$^{-1}$), obtained by 
folding of the differential cross section in [16] with the simplified 
reactor spectrum of [25] and applying quenching factors from [22]. 
The tradeoff between endpoint energy and 
rate with increasing atomic mass is evident. The 
instrumental resolution smear is not included (this should 
spill the signal toward higher e.e.e.'s). 
{\it Inset}: Similar signal expected 
from a high-purity filtered (Fe+Al) neutron beam of 24 keV (2 keV 
FWHM). Other available energies are 2 keV (Sc+Ti), 55 keV (Si+S) and 144 
keV (Si+Ti).} 
\end{figure}

\vspace{1mm}
{\bf 3.3 Weakly Interacting Massive Particles.}

\vspace{1mm}
WIMPs constitute one of the best candidates for Dark Matter at the 
galactic level \cite{WIMPreview}. Numerous ongoing and planned searches 
aim at the detection of keV nuclear recoils induced by WIMP elastic 
scattering \cite{dan}. The difficulty there is the scarce number of interactions 
expected ($<$1-10 recoils/kg target mass/day) and hence the need for 
background reduction and/or rejection. Besides, the WIMP signal 
monotonically decreases with increasing energy, favoring the lowest 
possible detector threshold. The advantage of gas detectors with good space 
resolution was recognized early on \cite{jim}; nuclear recoils can be 
distinguished from competing backgrounds by their much shorter trajectories 
(\frenchspacing{Fig. 4}), with near perfect rejection ability. Unfortunately, 
the few prototypes built so far \cite{tpc,martoff} 
(based on a MWPC design) are 
limited by modest spatial resolutions of O(1)mm,
the possibility of spurious pulses at low energies
and the need to operate at reduced pressure to obtain a particle 
identification ability in the keV region. As a result of the last, their volume-to-target 
mass is less than optimal (multi-m$^{3}$/kg) in view of the low 
signal rate expected. The radically superior space resolution of $\mu$MS can 
overcome this setback, reverting this ratio (multi-kg/m$^{3}$). 
Other advantages can be listed: 

\vspace{1mm}
$\bullet$ High-spatial resolution gaseous detectors can determine the initial 
direction of a recoiling nucleus. The Sun's movement through the galactic 
halo produces a pseudocollimation of the WIMP flux, which in turn engenders 
an anisotropy in recoil direction \cite{spergel}. This gives rise to a wickedly 
complex, geographically-dependent WIMP signature. Not only this 
anisotropy changes with a diurnal periodicity due to the Earth's rotation, but 
it goes out of phase yearly (due the Earth's orbital motion). Its observation 
would constitute a far more convincing evidence for WIMP dark matter than 
the so-called ``annual modulation'', in face of a number of seasonal effects 
able to mimic the last. 

\vspace{1mm}
$\bullet$ Fluorine is the best target for the detection of spin-dependent WIMP 
interactions \cite{john2}. Modest-mass CF$_{4}$-filled $\mu$MS can explore a 
substantial fraction of supersymmetric WIMP parameter space beyond the reach of the 
most ambitious cryogenic proposals \cite{njp}. Xe loading of the same 
device would be optimal 
for spin-independent (coherent) interactions due to its large nuclear 
mass. CIF$_{3}$ and CBrF$_{3}$ (non-toxic, non-flammable inexpensive gases used in fire 
extinguishers) offer the best of both worlds but are yet to be tested in 
a $\mu$MS.

\vspace{1mm}
$\bullet$ A putative population of solar-bound WIMPs, gravitationally captured via several 
plausible mechanisms is only within reach of detectors having sub-keV 
ionization thresholds \cite{myreview}. Again, the combination of mass and threshold 
requirements has made this search impossible until now. 

\vspace{1mm}
$\bullet$ The competitiveness of compact, direction-sensitive gaseous 
WIMP detectors with respect to 
tonne or multi-tonne solid-state devices has been emphasized before \cite{martoff}. 
\begin{figure}[tbp]
\epsfxsize = \hsize \epsfbox{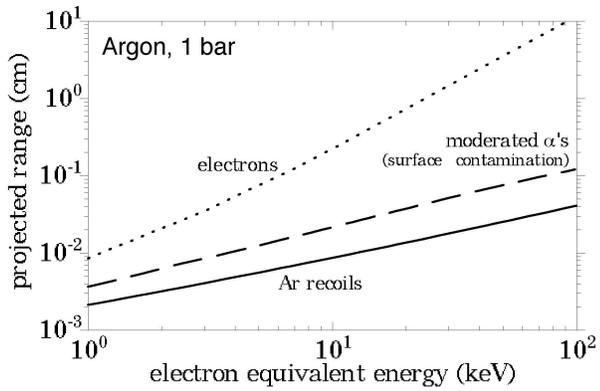}
\caption{Few-keV WIMP-induced nuclear recoils can be efficiently separated from 
competing backgrounds in high spatial resolution 
gas detectors due to their characteristic shorter trajectories.} 
\end{figure}

\vspace{1mm}
{\bf 4. Conclusions}

\vspace{1mm}
The feasibility of other applications is still under study. For instance, in beta-decay 
kinematic tests for neutrino mass, the achievable sensitivity 
depends similarly on source intensity and energy resolution of the 
spectrometer. The fast response of a tritium-spiked $\mu$MS may allow 
to use (before pile-up constrains become an issue) a decay rate amply able to 
compensate for a diminished energy resolution vis-a-vis dedicated spectrometers. 
An alluring possibility is the ``miniaturization'' of neutrino oscillation 
experiments by use of a compact-yet-intense $^{3}$H source like that described in 
\cite{3h}, surrounded by an array of $\mu$MS chambers. The 
reachable $\overline{\nu}_{e}\to \nu_{x}$
parameter space 
is largely unexplored and may, for a large enough source, extend to the 
region favored to explain the solar neutrino deficit  
(as a reference, a preliminary calculation shows that a 10 kg $^{3}$H 
source surrounded by 10 m of $\mu$MS-instrumented Xe at 1 atm can 
bring the  $\overline{\nu}_{e}\to \overline{\nu}_{\mu}$ sensitivity 
down to the $\Delta m^{2}\sim 10^{-4}$ level). The same 
set-up could be simultaneously used  for a high-sensitivity neutrino magnetic 
moment search, where a departure from the Standard Model 
prediction in the low-energy antineutrino-electron scattering cross section  
would be sought\cite{munu}. The 
neutrino flux from a realistic $^{3}$H source can be a factor $>$100 larger than 
in reactor experiments but the maximum electron recoil energy is $\sim$1 keV, again 
severely constraining the choice of detector \cite{neganov} and making 
of $\mu$MS an ideal choice. 

All in all, we foresee a busy future for MICROMEGAS in the field of 
low-background, low-energy experiments.

\vspace{1mm}
{\bf 5. Acknowledgments:}

\vspace{1mm}
We are indebted to J. Derre and P. Gorodetzky for many useful 
discussions and suggestions.

\end{document}